\documentstyle[preprint,prl,aps]{revtex}

\begin{document}
\draft

\title{
Equivalence of the antiferromagnetic Heisenberg ladder to a
single $S=1$ chain
}
\author{ Steven R.\ White}
\address{
Department of Physics,
University of California,
Irvine, CA 92717
}
\date{\today}
\maketitle
\begin{abstract}
I introduce two continuous transformations between the $S=1$ Heisenberg
chain and the antiferromagnetic $S=1/2$ Heisenberg ladder.  Both
transformations couple diagonally situated {\it next nearest neighbor}
$S=1/2$'s to form each $S=1$.  Using the density matrix
renormalization group, I demonstrate that the two systems are in the
same phase.  Furthermore, I find that the hidden topological
long-range order characterizing the $S=1$ system is even stronger in
the isotropic two-chain system.
\end{abstract}


\narrowtext
In the dozen years since Haldane's conjecture\cite{haldane} that
antiferromagnetic Heisenberg chains with integral spin are gapped,
while half-integral spin chains are gapless, our understanding of these
systems has increased tremendously. New analytical approaches,
exactly soluble models, experimental systems and techniques, and numerical
techniques have provided convincing evidence in support of the
conjecture\cite{haldanereview}.  One of the most instructive developments
was the discovery of the AKLT model\cite{AKLT},
an exactly soluble, gapped $S=1$ chain system, differing from the Heisenberg
system only by the addition to the Hamiltonian of a biquadratic term
$-\frac{1}{3}( {\hbox{\bf S}}_{i} \cdot {\hbox{\bf S}}_{j})^2$.
The AKLT state is now believed to be an ideal example of the ``Haldane''
state of the $S=1$ system. It has a hidden form of topological long-range
order\cite{dennijs,girvin}, measured by a ``string'' correlation function,
which has been determined to be present in the Heisenberg $S=1$
chain\cite{girvin,whitehuse}.

More recently, attention has been focused on the problem of the
antiferromagnetic Heisenberg ladder (AFHL), two antiferromagnetically
coupled antiferromagnetic $S=1/2$
chains\cite{dagotto,strong,barnes,rice}.
There is now evidence that
this system has a gap for all nonzero interchain couplings
$J_\perp$\cite{barnes}.
It has been suggested that the AFHL system and the $S=1$ Heisenberg system
may be in the same phase\cite{privatecoms}.
I consider two systems to be in the same phase if there is a
continuous path through model parameter space from one system to the
other, without crossing any phase boundaries or critical points, and
consequently without change in any broken symmetries or disappearance or
appearance of gaps.
The {\it ferromagnetically coupled} Heisenberg ladder (FHL),
for sufficiently strong interchain coupling, has been known for
some time to be in the same phase as a single $S=1$ chain;  in the limit
$J_\perp\to-\infty$, the two models are identical.
For the AFHL case,
the origin of the gap is clear in the large $J_\perp$ limit, where a single
``rung'' of the ladder has a gap of size $J_\perp$ between the singlet
and triplet states. This origin for the gap seems completely different from
the origin in the Haldane case, and it is natural to assume that there are
two distinct phases.  Furthermore, the most obvious path connecting the
systems, varying $J_\perp$ from
positive to negative values, passes through the gapless point $J_\perp=0$.
Xian has proposed\cite{xian} that for a region of large $J_\perp$, the
system is in a dimerized phase consisting primarily of singlets on each
rung, while for smaller $J_\perp > 0$, the system is either
in a Haldane phase or in another gapped phase which has no topological
long-range order.
A useful example of such a gapped, orderless state has
been thought to be the dimer resonating valence bond (RVB)
state\cite{kivelson},
which has been used to describe the qualitative features of the AFHL
system\cite{rvbprl,rice}. However, in the RVB picture there is no distinction
between the rung-singlet dimerized phase and the gapped, orderless phase.

In this letter I show that the AFHL system does belong to
the same phase as the $S=1$ chain, the Haldane phase, for {\it all}
values of $J_\perp > 0$. The dimerized phase, the Haldane phase,
and the dimer RVB phase are all {\it identical}.
These surprising results are possible because, unlike the FHL case,
diagonally-situated next-nearest neighbor spins couple to form an effective
$S=1$, rather than two spins on the same rung. The AFHL and FHL systems
belong to the same phase in a slightly more limited sense, in that a shift
of one chain relative to the other by one lattice spacing is necessary in
constructing the path connecting the systems.  I demonstrate these results
by constructing explicit paths, and calculating the properties of the system
to high accuracy, as the parameters are varied,
using the density matrix renormalization group
(DMRG)\cite{white}.
In addition to calculating the gap, I calculate the limiting value
of the string correlation function. Surprisingly, the hidden topological
order is stronger in two isotropically coupled chains than in the $S=1$ chain.
Furthermore, I show that the dimer RVB state on two coupled chains
has ``perfect'' topological order, just like the AKLT state.
In fact, in the composite spin model\cite{composite},
which can be thought of as both a single chain $S=1$ system and
a $S=1/2$ ladder, the dimer RVB state {\it is}  the AKLT state.

I consider the Heisenberg Hamiltonian
\begin{equation}
H = \sum_{ i,j } J_{ij}
	{\hbox{\bf S}}_{i} \cdot {\hbox{\bf S}}_{j}.
\end{equation}
Figure 1 illustrates the various models considered. In all
cases, the intrachain coupling is taken as $J_{ij} = J=1$, while additional
interchain couplings are as shown. The mapping used for the FHL system,
which has been studied in some detail\cite{watanabe}, is shown in Fig. 1(a).
Figure 1(b) shows a mapping for the AFHL case. Here next-nearest neighbor
spins, which because of the antiferromagnetic order tend to be in a triplet
state, are grouped in pairs to form $S=1$ spins.
For $J_2=0$, we have the AFHL system.
In the limit $J_2\to -\infty$, the singlet states of the spins coupled
by $J_2$ are pushed to $\infty$, and the system is identical to an
$S=1$ single chain, with effective coupling $J_{\rm eff} = 3/4J$.

Figure 2 shows the evolution of the gap as $J_2$ is varied for the system
shown in Fig. 1b. The gap is plotted as a function of $x_0^{1/2}$,
where $x_0$ is the probability (and $x_0^{1/2}$ the amplitude)
that a pair of spins $i$,$j$ coupled by
$J_2$ are in a singlet state,
$x_0 = 1/4 - \langle {\hbox{\bf S}}_{i} \cdot {\hbox{\bf S}}_{j}\rangle$.
The results were obtained by extrapolating from three system sizes,
$L=19,31,39$, using open boundary conditions. The extrapolation
used a polynomial fit in $1/L^n$, with the $1/L$ term excluded.
The finite system version of DMRG was used, keeping 60 states, with a
typical discarded weight of $2 \times 10^{-6}$.
The point at $x_0=0$ is the taken from previous results for the $S=1$
chain, $\Delta \cong 0.41050(2) \times 3/4$. The line is a fourth-order
polynomial fit to the data. The typical deviation of the points from the
fit is $\cong 10^{-4}$. At $J_2=0$, the probability of finding a
diagonally situated pair of spins in a triplet state is $96.2\%$,
a surprisingly high number reflecting the short-range
antiferromagnetic order, indicating that even at $J_2=0$, the system is not
too far (in this sense) from the $S=1$ system.

Figure 1(c) shows another mapping between the AFHL system and
a single $S=1$ chain. In this case a {\it finite}  antiferromagnetic
coupling $J_3=1$ turns two chains into the composite spin model shown in
Fig. 1(d), if one shifts the upper chain to the left by one spacing. The
composite spin model is identical to a $S=1$ chain, except for some extra
excited states involving singlet modes of a rung. The total spin of
each rung commutes with the Hamiltonian, so the eigenstates can all
be classified by the total spin on each rung. The set of eigenstates
with no singlet modes on any rungs corresponds to the spectrum of
the $S=1$ Heisenberg chain.

Figure 3 shows the gap as $J_3$ is varied from $0$ to $1$.
In this case the results for finite $L$ as well as the extrapolation
to $L \to \infty$ are shown. To demonstrate conclusively the robustness
of this mapping, large systems were used (up to $L=100$).
Again the finite system version of DMRG was used, this time keeping
up to 100 states, for a typical discarded weight of $10^{-8}$.
The data was fit very well with a 14 parameter polynomial function with
terms of the form $J_3^m/L^n$, excluding $n=1$. The resulting gap for
$L \to \infty$ as a function of $J_3$, accurate to four or five digits, is
\begin{equation}
\Delta =
0.50249-0.227786J_3 + 0.074252 J_3^2 + 0.067215 J_3^3 - 0.005681 J_3^4
\end{equation}
This very smooth evolution of the gap shows that no phase transitions
of either first or second order occur along this path.

Given these results, the AFHL system must exhibit the same
topological order known to exist for Haldane chains.
This broken symmetry is measured by the string correlation function
\cite{dennijs,xian}
\begin{eqnarray}
g(\ell) = \langle S^z_0 (\prod_{k=1}^{\ell-1}e^{i\pi S^z_k}) S^z_\ell
\rangle .
\label{string}
\end{eqnarray}
For coupled chains, the expression for $g(\ell)$ is the same as for a
single $S=1$ chain if we take
\begin{eqnarray}
\hbox{\bf S}_k=\hbox{\bf S}_{k,1}+\hbox{\bf S}_{k,2} ,
\label{Sdef}
\end{eqnarray}
where the indices 1, 2 indicate the two $S=1/2$ spins which we expect to
combine to form a single effective $S=1$ spin.  If sites 1 and 2 are
taken from the same rung, as
would be appropriate for the FHL system, the string correlation function
decays very rapidly to zero, with a decay length of about 1.   Figure 4
shows $g(\infty)$ when sites 1 and 2 are next-nearest neighbors, as
shown in Fig. 1(c), as $J_3$ is varied from $0$ to $1$.  As many as 108
states were kept in the calculations, for which no finite-size extrapolation
is necessary. Details of the procedure are described in Ref.
\cite{whitehuse}. The result at $J_3=0$, $g(\infty) = -0.38010765$ is {\it
larger}  in magnitude than the result for the $S=1$ chain
($J_3=1$)\cite{whitehuse}, $g(\infty) = -0.374325096(2)$.

I have also calculated $g(\infty)$ as a function of $J_\perp$ with
$J_3=0$. Fig. 5 shows the results, plotted as a function of
$J_\perp/(1+J_\perp)$. Near $J_\perp=0$, I find $g(\infty) \sim
J_\perp^{1/2}$. At the maximum point shown,  $J_\perp=1.3$,
$g(\infty) = -0.387263374$. At $J_\perp=\infty$, I find $g(\infty) = 1/4$.

The behavior of $g(\infty)$ at $J_\perp=\infty$ is easily understood.   In
this limit, the ground state consists of singlets on each rung. If sites $i$
and $j$ are part of such a singlet, then necessarily $S^z_i+S^z_j=0$. Using
this, all but a few of the $e^{i\pi S^z_k}$ terms in (3) cancel, leaving a
factor of $-i/2$ for each of the two ends. Hence $g(\infty) = -1/4$ at
$J_\perp=\infty$, in agreement with our results. It is useful also to
consider a ``normalized'' string correlation function, defined by
\begin{eqnarray}
\tilde g(\ell) = \frac{-g(\ell)}
	{\langle (S^z_{k,1}+S^z_{k,2})^2 \rangle^2} ,
\label{gstring}
\end{eqnarray}
where $k$ is any $S=1$ ``site''.
In $\tilde g(\ell)$, one obtains contributions only from those spin
configurations for which $S^z_{k,1}+S^z_{k,2}=\pm1$ for the endpoints $k=0$
and $k=\ell$.  The normalized function shows more clearly whether there are
defects in the string order {\it between}  sites $0$ and $\ell$. For
$J_\perp=\infty$, $\tilde g(\infty) = 1$, indicating perfect order within
the string. For a $S=1$ single chain, the denominator of (5) is $4/9$, and
the AKLT model also has perfect order, $\tilde g(\infty) = 1$. For the
Heisenberg $S=1$ chain $\tilde g(\infty) = 0.84$. For the AFHL
system with $J_\perp=1$, $\tilde g(\infty) = 0.924$.

The rung-singlet ground state of $J_\perp=\infty$ can be considered a limiting
case of a more general set of wavefunctions, dimer RVB states, which are
themselves limiting cases of the set of short-range RVB states.
A dimer RVB state also has perfect string order. The proof is
straightforward and similar to that of the strong coupling case:
pairs of spins which are parts of singlets cancel in their effect on
$\tilde g(\ell)$. Valence bond configurations for which the contribution
to $\tilde g(\ell)$ is not 1 are either ``staggered''
configurations\cite{rvbprl},
which are neglible in the thermodynamic limit, or they have at least one
long-range (non-dimer) bond, with exactly one end within the region $0$ and
$\ell$.  This means that the confinement of long range bonds within the RVB
picture\cite{rvbprl}
is directly measured by the string correlation function.
Probably any short-range RVB state has nonzero string order.
Note also that
staggered configurations in the two-chain system corresponds to spin-Peierls
order in the $S=1$ chain, which has been known to be forbidden \cite{arovas}.

The AKLT state of the $S=1$ chain is constructed within the composite
spin model (Fig. 1(d)) by first making intrachain near-neighbor
singlets so that each rung has
two singlets attached, one to the left and one to the right. Then,
one symmetrizes the spins on each rung. It has apparently not been noticed
before that this state is also the dimer RVB state for the composite
spin model, provided that one eliminates the two spin-Peierls valence
bond configurations, which are neglible in the thermodynamic limit.
(Also, no single-rung dimers are allowed.)
The symmetrizing operation corresponds to summing over all dimer valence
bond configurations, which have an especially simple structure in the
composite spin model. Note that the lattice of the composite spin model
is symmetrical with respect to interchange of the sites on any rung.
Consequently, there
is only one reasonable dimer state that can be constructed, as opposed to
the AFHL system, for which an infinite number of dimer states
can be constructed, corresponding to different amplitudes
for horizontal and vertical bonds and to different correlations between
horizontal and vertical bonds.

The mappings discussed here also explain the relationship between the
$S=1/2$ end states seen at open ends of $S=1$
chains\cite{kennedy,hagiwara,white} and later also on AFHL
systems with an extra site on the end of one of the chains\cite{rvbprl}.
Because of the shift of one chain relative to the other implicit in the
transformation of one system to the other, an open $S=1$ end is equivalent
to a ladder with an extra site on one chain.

An important consequence of these results concerns doped ladders and chains.
Recent studies of Hubbard and $t$-$J$ ladders have found evidence for
a spin-liquid phase with strong pairing correlations for moderate
doping\cite{noack}. It may be possible to dope ladder compounds to form
a physical analog of such a system. Our results suggest that doped
Haldane chains may also exhibit a spin-liquid phase with strong pairing
correlations.

After this work was finished, I received a preprint by H.
Watanabe which comes to some of the same conclusions reached here.
I thank Ian Affleck for helpful conversations, and Y. Xian and H. Watanabe
for sending preprints of their work.
I acknowledge support from the Office of Naval
Research under grant No. N00014-91-J-1143.
This work was supported  in part by the University of California
through an allocation of computer time.

\begin{figure}
\caption{Several mappings between two coupled $S=\frac{1}{2}$
Heisenberg chains and a single $S=1$ chain.
}
\label{a}
\end{figure}
\begin{figure}
\caption{Gap between the ground state and first excited state of
the ladder system shown in Fig. 1(b) as $J_2$ is varied, plotted
as a function of $x_0^{1/2}$,
where $x_0 = 1/4 - \langle {\hbox{\bf S}}_{i} \cdot {\hbox{\bf
S}}_{j}\rangle$, with $i$ and $j$ coupled by $J_2$.
}
\label{b}
\end{figure}
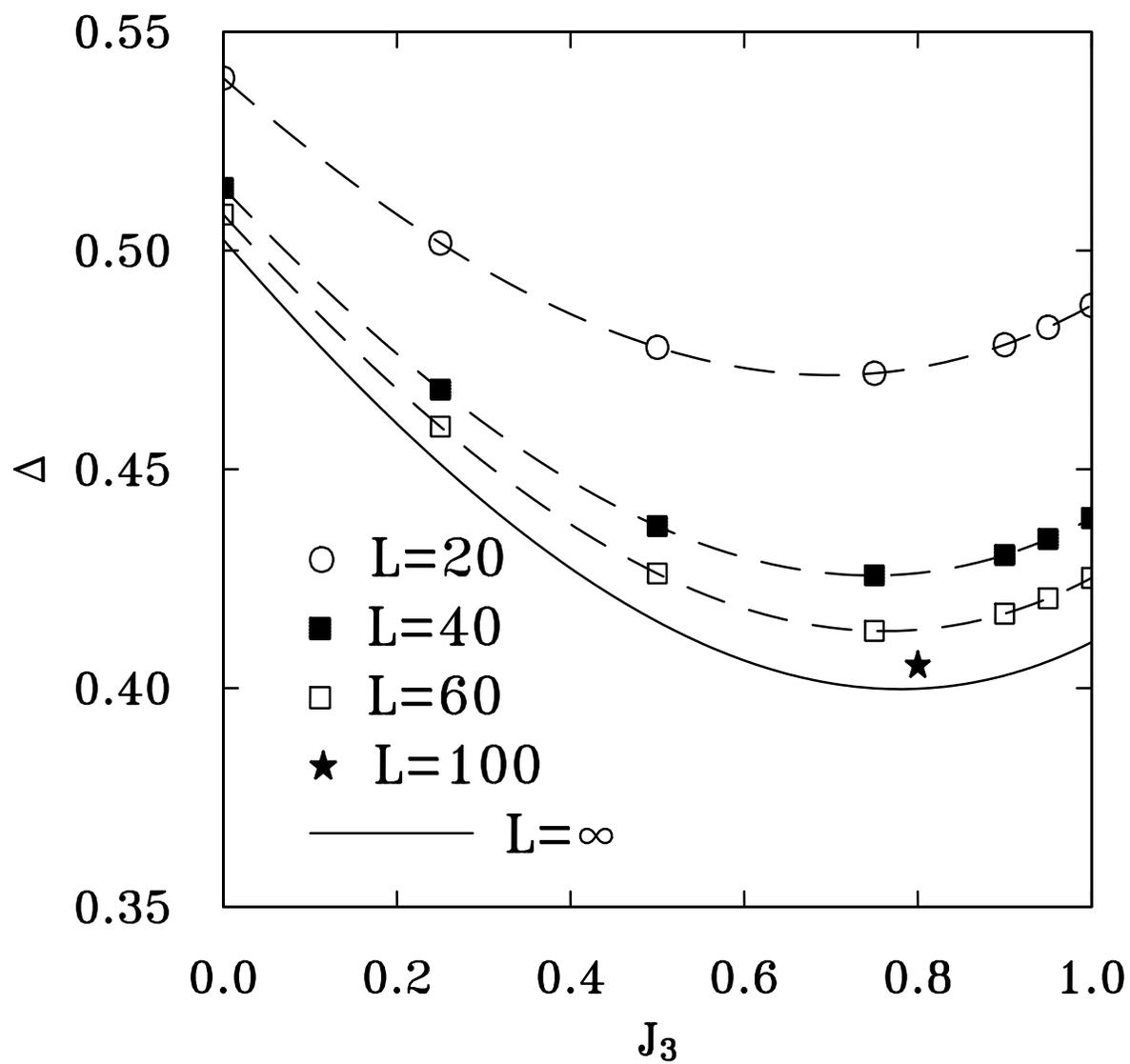
\begin{figure}
\caption{Gap as a function of $J_3$ for the ladder system shown in Fig.  1(c).
}
\label{c}
\end{figure}
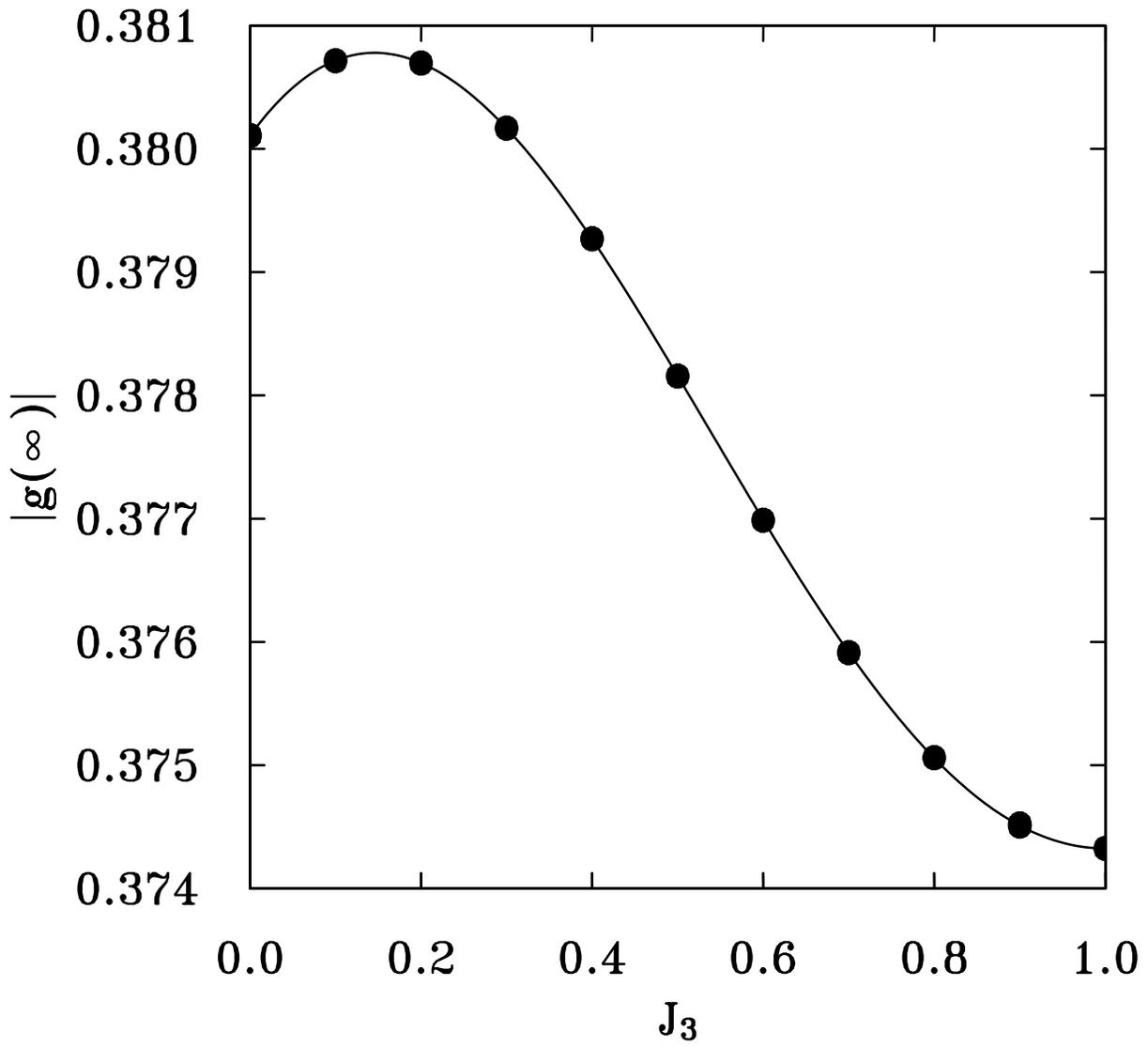
\begin{figure}
\caption{Limiting value of the string correlation function as a function
of $J_3$ for the system of Fig. 1(c).
}
\label{d}
\end{figure}
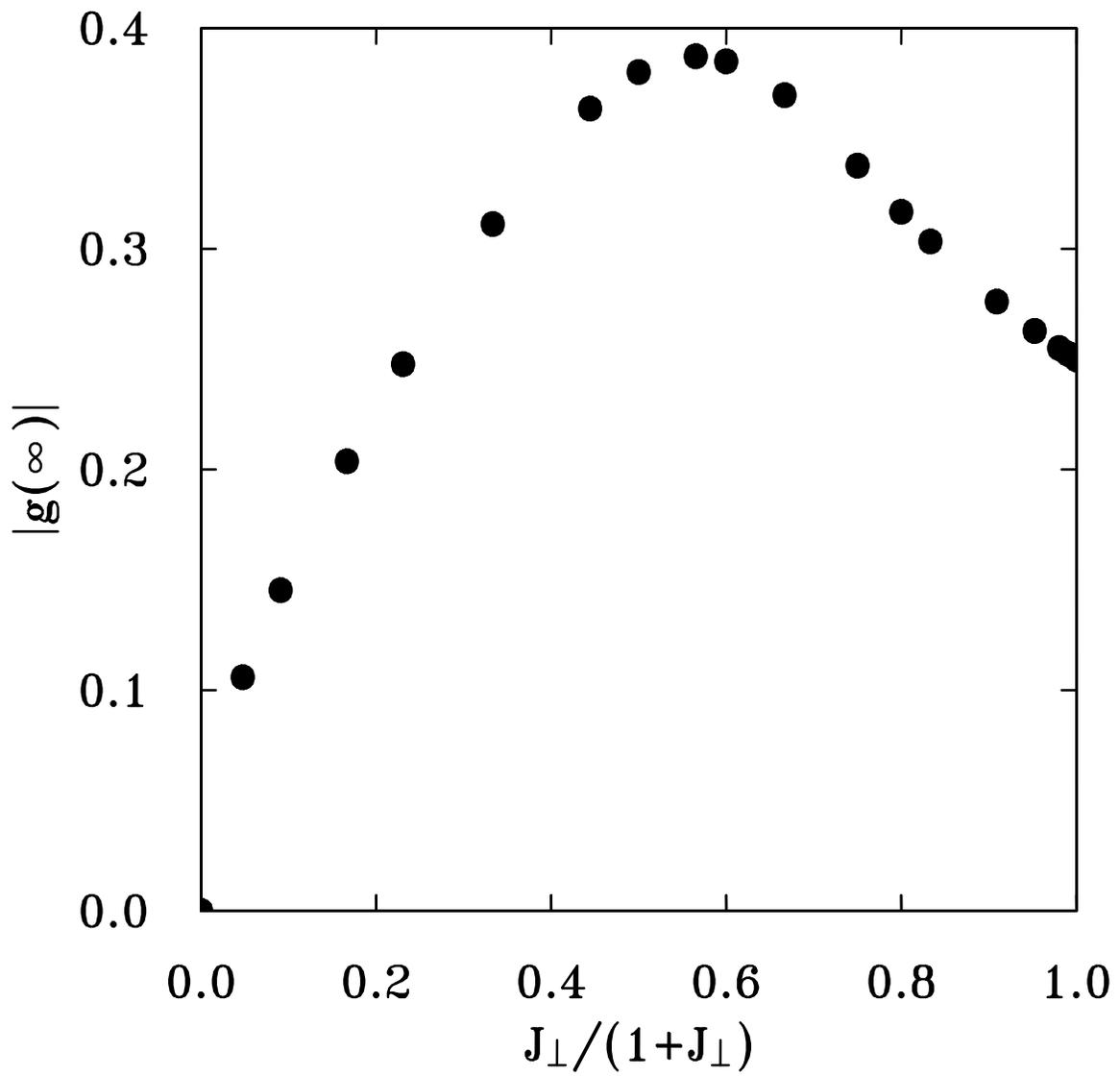
\begin{figure}
\caption{Limiting value of the string correlation function as a function
of $J_\perp$ for the AFHL system (with $J_2=J_3=0$).
}
\label{e}
\end{figure}

\end{document}